\documentclass[runningheads]{llncs}
\usepackage[T1]{fontenc}
\usepackage{graphicx}
\usepackage{amsmath,amssymb,amsfonts}
\usepackage{orcidlink}
\usepackage{listings}
\usepackage{enumitem}
\usepackage{adjustbox}   
\usepackage{tabularx}    
\usepackage{array}       
\usepackage{fancyvrb}

\begin{document}
\title{Robust DDoS‑Attack Classification With 3D CNNs Against Adversarial Methods}
\titlerunning{Robust DDoS‑Attack Classification With 3D CNNs}
%
\author{Landon~Bragg \and 
Nathan~Dorsey \and 
Josh~Prior \and 
John~Ajit \and 
Ben~Kim \and 
Nate~Willis \and 
Pablo Rivas\orcidlink{0000-0002-8690-0987}
}
\authorrunning{L. Bragg et al.}
\institute{School of Engineering and Computer Science, Department of Computer Science, Baylor University, Waco, TX 76798, USA\\ 
\email{\{Landon\_Bragg1, Nathan\_Dorsey1, Josh\_Prior1, John\_Ajit1, Ben\_Kim1, Nate\_Willis1, 
Pablo\_Rivas\}@Baylor.edu}}
\maketitle              
\begin{abstract}
Distributed Denial-of-Service (DDoS) attacks remain a serious threat to online infrastructure, often bypassing detection by altering traffic in subtle ways. We present a method using hive-plot sequences of network data and a 3D convolutional neural network (3D CNN) to classify DDoS traffic with high accuracy. Our system relies on three main ideas: (1) using spatio-temporal hive-plot encodings to set a pattern-recognition baseline, (2) applying adversarial training with FGSM and PGD alongside spatial noise and image shifts, and (3) analyzing frame-wise predictions to find early signals. On a benchmark dataset, our method lifts adversarial accuracy from 50--55\% to over 93\% while maintaining clean-sample performance. Frames 3--4 offer strong predictive signals, showing early-stage classification is possible.
\end{abstract}

\section{Introduction}\label{sec:intro}

Distributed denial-of-service (DDoS) attacks remain a critical challenge for network infrastructure, as adversaries increasingly use subtle changes in traffic patterns to bypass conventional detection systems. These attacks, which overwhelm services with traffic to cause disruption, require new methods that can model both the structure and timing of traffic flows.

Recent progress in spatio-temporal representation learning has shown strong results in related areas like video classification, environmental prediction, and graph modeling. 3D convolutional neural networks (3D CNNs) and related architectures have been used to extract spatial and temporal patterns from structured input \cite{10.1109/iccv.2015.510,10.1109/iccv.2017.590,10.1007/978-3-319-46487-9_50,10.1109/cvpr.2016.573}. These ideas have been extended through self-supervised pretraining techniques \cite{10.1109/cvpr.2017.607,10.1109/cvpr.2019.00413,10.48550/arxiv.2006.11476} and applied to tasks involving flow graphs \cite{10.1007/11427834_1}, environmental data \cite{10.48550/arxiv.2007.11836}, lifecycle analysis \cite{10.1007/978-3-319-54660-5_54}. Surveys highlight the relevance of spatio-temporal models for structured data mining \cite{10.48550/arxiv.1906.04928,10.48550/arxiv.2108.11575}. Despite this progress, there is limited work on how these models respond to adversarial inputs in the context of network traffic classification. While some work calls for disentangled and interpretable representations to improve robustness \cite{10.48550/arxiv.2202.04821}, specific methods for adversarially hardened classification are still needed.

In this work, we propose a system that uses hive-plot sequences of network-flow data as input to a 3D CNN, allowing both spatial structure and temporal change to be modeled. We combine this with adversarial training using Fast Gradient Sign Method (FGSM) and Projected Gradient Descent (PGD), plus simple spatial augmentations (rotation, shear, zoom, and Gaussian noise). Frame-wise analysis of predictions shows that early detection is possible and highlights the most informative time steps. Together, these steps lead to a robust, GPU-accelerated pipeline for DDoS detection.

The rest of this paper is organized as follows. Section~\ref{sec:related} reviews related work. Our method is described in Section~\ref{sec:approach}, and experimental setup and results are given in Sections~\ref{sec:experiments}–\ref{sec:results}. Section~\ref{sec:analysis} provides robustness analysis and notes on deployment. We conclude in Section~\ref{sec:conclusion} with remarks on limitations and ethics. Code is available at \url{https://github.com/Landon-Bragg/DDoS_Attack_Classification}.

\section{Related Work}\label{sec:related}
Early detection systems for DDoS attacks relied on rule-based and threshold approaches to monitor aggregate traffic volumes or flow counts to flag suspicious network traffic. While implementing these defenses is simple, these methods can generate false positives due to legitimate spikes in traffic (such as flash crowds) and cannot adapt to changing attack patterns as they become more sophisticated. Machine learning techniques have been applied to network traffic features such as packet arrival times, byte counts, and header statistics in hopes of overcoming these limitations. Guarino et al.~\cite{Guarino} found that convolutional neural networks trained on raw traffic time series captured temporal correlations to better detection over traditional classification methods. While they made strides in this topic, these models remain vulnerable to adversarial perturbations that can subvert the decision boundaries learned by the model.

Graph-based visual methods offer multiple-dimensional views of the network traffic relationship flows into a spatial layout. Rivas et al.\cite{Rivas} found that representing flows as hive plots mapping features to axes and encoding edge weights by radial distance, allowed CNNs’ spatial feature extraction capabilities to shine. This representation improves interpretability and classification performance under difficult classifying conditions. While there are benefits, research in adversarial machine learning has shown that models trained solely on clean data can be entirely misled by minimal perturbations~\cite{goodfellow2015explainingharnessingadversarialexamples}. Techniques such as FGSM and PGD have been used to attack and defend image classifiers, but their application to network security has not yet been investigated. CNNs have also been proven to fail when faced with even minor augmentations as discussed in~\cite{Augmentation}. This indicates the need for a model that is capable of being robust against common practices that cause these models to fail. 

While research has begun on bringing adversarial training into machine learning-based detection systems, to our knowledge, no prior study has combined hive plot sequence representations with 3D CNN architectures and a mixed adversarial-augmented training protocol for DDoS classification. Our work bridges the gap of encoding temporal evolution via multi‐frame hive plot sequences, hardening a lightweight 3D CNN with both stochastic data augmentations and FGSM/PGD adversarial examples, and conducting frame‑wise robustness analysis to identify the most informative temporal windows for reliable detection.

\section{Approach}\label{sec:approach}
\subsection*{Problem Setup}

Prior work has been done towards the task of making a DDoS-attack classifier robust against adversarial augmentations and perturbations of hive-plot images of network traffic. In particular, prior work \cite{Bonaci} accomplishes this task using two different CNN architectures, ResNet34 and MobileNetv2, with success. Our methodology closely follows the adversarial robustness formulation outlined in~\cite{Guarino}; however, we adapt it to consider a novel approach using a 3D CNN to evaluate each sequence of hive-plot images as one observation. The hive plots are grouped into sequences of 8 images, from timesteps $t_0$–$t_7$, and represent network data collected over a given time period. We believe that the sequencing of this data, that is, the spatiotemporal features related to the onset velocity of the attack or the burst rhythm, for example, could have meaningful information related to the classification of whether the network traffic is representative of a DDoS attack or not.

\begin{figure}[h]
    \centering
    \includegraphics[height=0.9\textwidth]{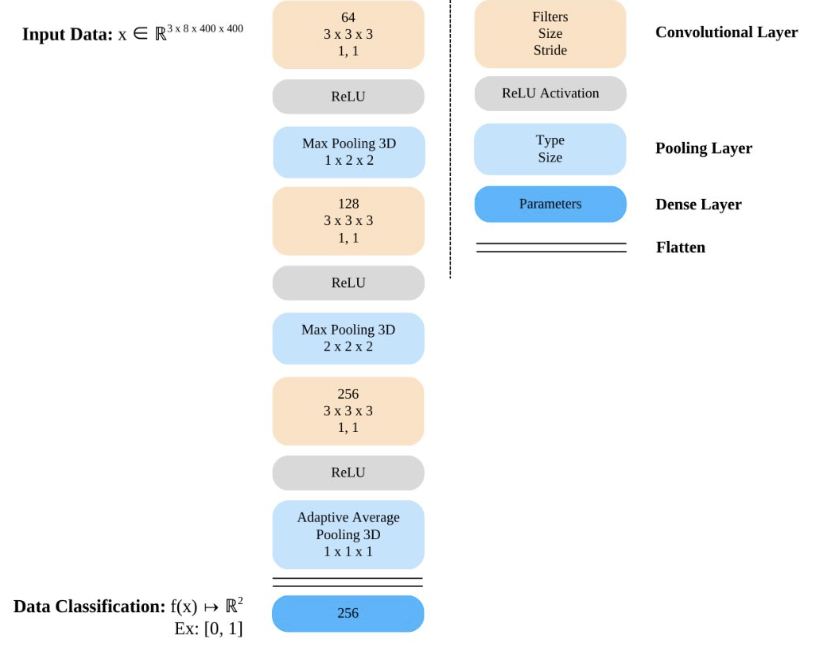}
    \caption{Model Architecture}
    \label{fig1:model}
\end{figure}

CNNs have long been around and utilized as the preferred model of computer vision tasks. 3D CNNs were first introduced in \cite{3DCNN} and have since become similarly significant in the context of computer vision for 3-dimensional objects and video recognition, such as in \cite{Tran}. 3D CNNs work similarly to a traditional 2D CNN, in our model, seen in Fig.~\ref{fig1:model}, we use three convolutional blocks, each including a convolutional layer, passing a filter over an input performing convolution, a ReLU activation layer to add non-linearity, a pooling layer, selecting the most important feature representations with pooling operations. 
Then we use a fully connected layer to achieve the binary classification. The 3D convolution operations themselves can be formulated as:
\[
Y_{c',\,d,\,h,\,w}
= \sum_{c=1}^{C_{\text{in}}}
\sum_{u=1}^{k_d}\sum_{v=1}^{k_h}\sum_{w'=1}^{k_w}
K_{c',c,u,v,w'}\;
X_{c,\,d+u-1,\,h+v-1,\,w+w'-1}.
\]
Here, \(Y_{c',d,h,w}\) is the output feature map at channel \(c'\), depth \(d\), height \(h\), and width \(w\). \(X\) is the input tensor with \(C_{\text{in}}\) channels, and \(K\) is the 3D convolutional kernel of size \(k_d \times k_h \times k_w\). The kernel slides over the input volume across depth, height, and width, computing a weighted sum at each location.

The distinguishing characteristic between a 3D CNN and a traditional 2D CNN like those used in \cite{Guarino} is the shape of the input and of the kernel used to perform convolution. In a 3D CNN, a 3D kernel is passed over a 3D input, capturing spatial patterns over not just 2 dimensions, height and width of some input matrix, but also depth (e.g., time, as in our case). This allows the model to learn spatiotemporal features by convolving over consecutive frames in video data, rather than only learning 2D features frame-by-frame, as in a traditional CNN. We consider applying this concept to our image sequences, treating each sequence of 8 frames as a single observation, that is, one instance of network traffic collected over some point in time. See the appendix for additional details.

Following \cite{Guarino}, we frame the adversarial robustness training as a saddle point optimization problem, solving an inner maximization problem and an outer minimization problem. That is, the adversary, the adversarial attacks and perturbations, attempts to maximize the loss of the classifier model while the model seeks to find the set of model parameters $\theta$ which minimize the overall loss despite the best efforts of the adversary.

In choosing adversarial attacks to use in this work, we looked to \cite{Guarino} as well as considered which augmentations and perturbations were both well-established in the literature and would help our model achieve robustness against a wide variety of attacks. Similarly to \cite{Guarino}, we consider gradient-based attacks and gradient-free attacks.

\subsection*{Gradient-Based Attacks}

In gradient-based attacks, the attacker has access to the model they are attacking. In other words, a gradient-based attack leverages the gradient of some machine learning algorithm which uses a gradient for learning. We consider: Gradient Sign Attack, aka Fast Gradient Sign Method, or FGSM \cite{goodfellow2015explainingharnessingadversarialexamples} and 
Projected Gradient Descent Attack, or PGD \cite{MadryTowardsDLModelsResistant}

\paragraph{Gradient Sign Attack.}
The FGSM attack uses the gradient of the model to calculate the direction in which to modify the data to maximize the loss with respect to the input. It takes in the original input, a small constant $\epsilon$ controlling the magnitude of the perturbation, and the gradient of the loss function, using the sign of the gradient of the loss function to calculate the direction.

We can formulate it using a standard adversarial attack definition, and think of this adversary as figuring out how it can modify the data to best fool the model by finding a measure of each pixel’s contribution to the loss and perturbing it to maximize that loss. This method is a one-step attack, that is, it is not iterative and considers only the current example to construct the perturbation.

\paragraph{Projected Gradient Descent Attack.}

PGD is a similar gradient-based attack we considered for the perturbation of our data. Similarly, it uses the gradient of the model to perturb the data in a way which maximizes the loss of the classifier. However, unlike FGSM, it is iterative. At each iteration, it computes the gradient of the loss function with respect to the input and produces a small perturbation in the direction that maximizes the loss of the gradient. This allows it to create subtler and more effective perturbations than FGSM. We define its formal definition in the appendix. 

\subsection*{Gradient-Free Attacks}

The gradient-free “black box” attacks we used simply added augmentations to the data, without knowledge of the model architectures or parameters. Such augmentations considered in this work were rotation, noise, shear, and crop. Our values are listed later in the paper.

These attacks may not be the most sophisticated, but they are easy and quick to carry out, meaning that an attacker without access to the model and without much time, either in runtime or development time, can easily carry out such attacks, or similar, which augment the data significantly enough to cause it to fail.

\subsection*{Data Loader}

To efficiently process and structure the input data, we implement a custom data loader using PyTorch. Each sequence of \texttt{.png} images, representing an 8-frame hive-plot series, is converted into a 4D tensor with shape \((B \times C \times D \times H \times W)\), where \(B\) is the batch size, \(C\) the number of channels, \(D\) the sequence depth (i.e., number of frames), and \(H \times W\) the spatial resolution. These tensors are stored as \texttt{.pt} files, allowing for efficient storage and quick retrieval during training and evaluation. This design treats each image sequence as a unified spatiotemporal observation, aligning the task more closely with video classification problems such as those studied in \cite{Tran}, rather than treating individual frames in isolation. The resulting input tensors are well-suited for 3D convolutions, enabling the model to extract features jointly across depth, height, and width.

\subsection*{Training Regimes and Evaluation Framework}

We compare two training regimes: one based solely on clean data, and another incorporating adversarial examples. In the clean training setup, the model is trained using only unperturbed input sequences. Evaluation is conducted across four test conditions: clean inputs, spatially augmented data, and adversarial examples generated using both FGSM and PGD. For each condition, we report standard classification metrics, including accuracy, precision, recall, and the area under the ROC curve (AUC). In the adversarial training setup, the model is trained on a mixture of clean and perturbed sequences. This regime is designed to increase robustness against adversarial manipulations while maintaining accuracy on unperturbed samples. The same evaluation procedure is used, enabling a consistent comparison of model performance under both regimes.

\section{Experiments}\label{sec:experiments}
\subsection{Data}

The dataset used in this work was developed at Marist College and consists of 16,000 hive plot images representing simulated network traffic. The traffic includes both normal and attack scenarios, generated by issuing HTTP requests, either legitimate or malicious DDoS, to a honeypot designed to mimic a REST API endpoint. Each image is a hive plot with three axes: the leftmost axis encodes time since the start of capture (earlier packets appear closer to the origin), the vertical axis represents the country associated with the IP address, and the rightmost axis shows the source IP address. Lines connecting the axes have 50\% transparency, such that denser traffic regions appear darker due to overlap.

\begin{figure}[h]
    \centering
    \begin{tabular}{cccc}
        $t_0$ & $t_1$ & $t_2$ & $t_3$ \\ 
         \includegraphics[width=0.24\textwidth]{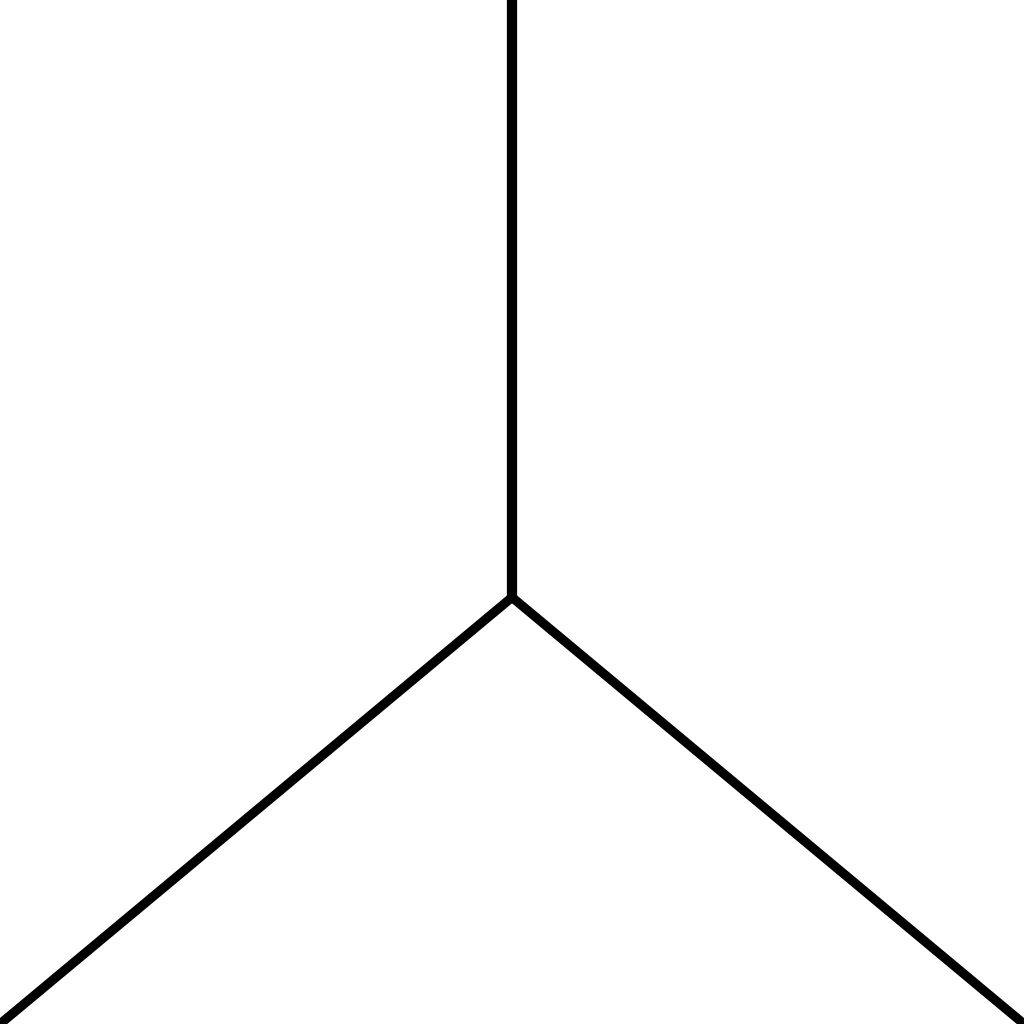} &  
         \includegraphics[width=0.24\textwidth]{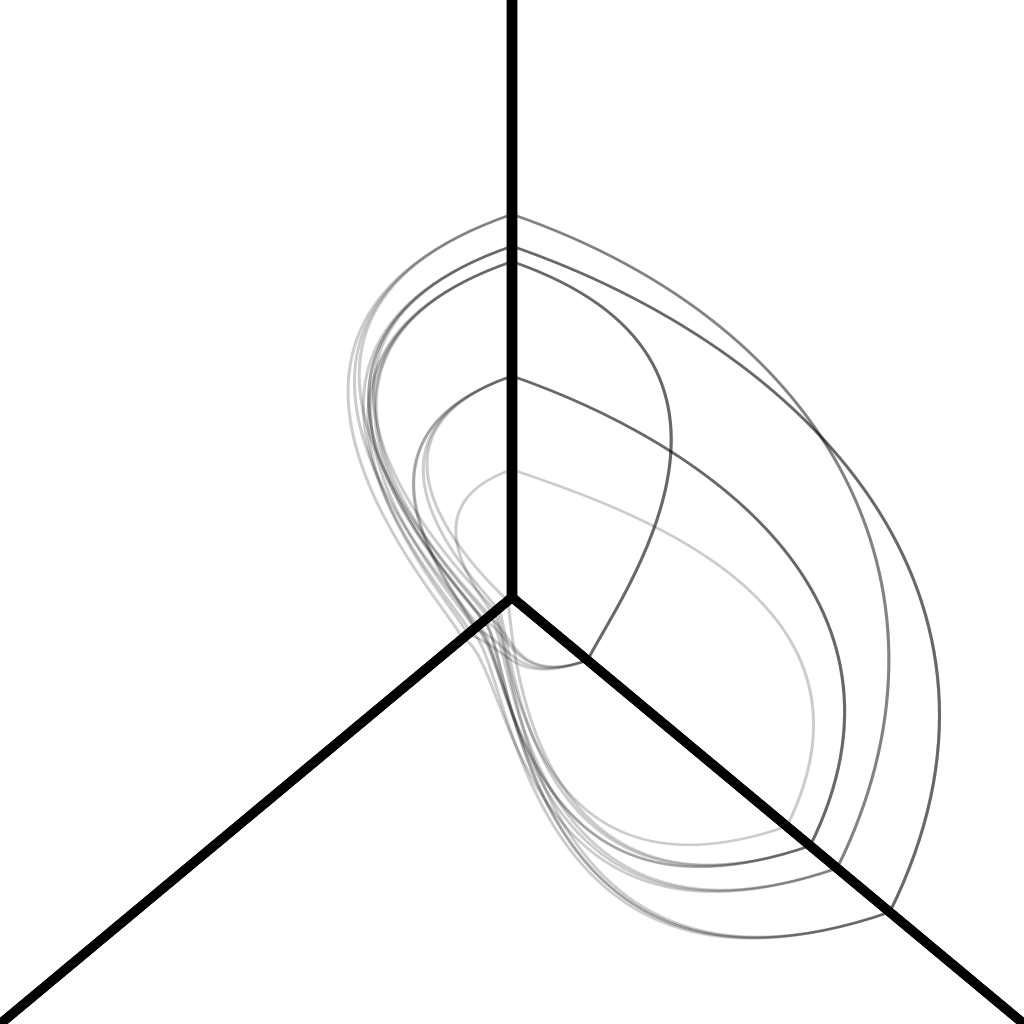} & 
         \includegraphics[width=0.24\textwidth]{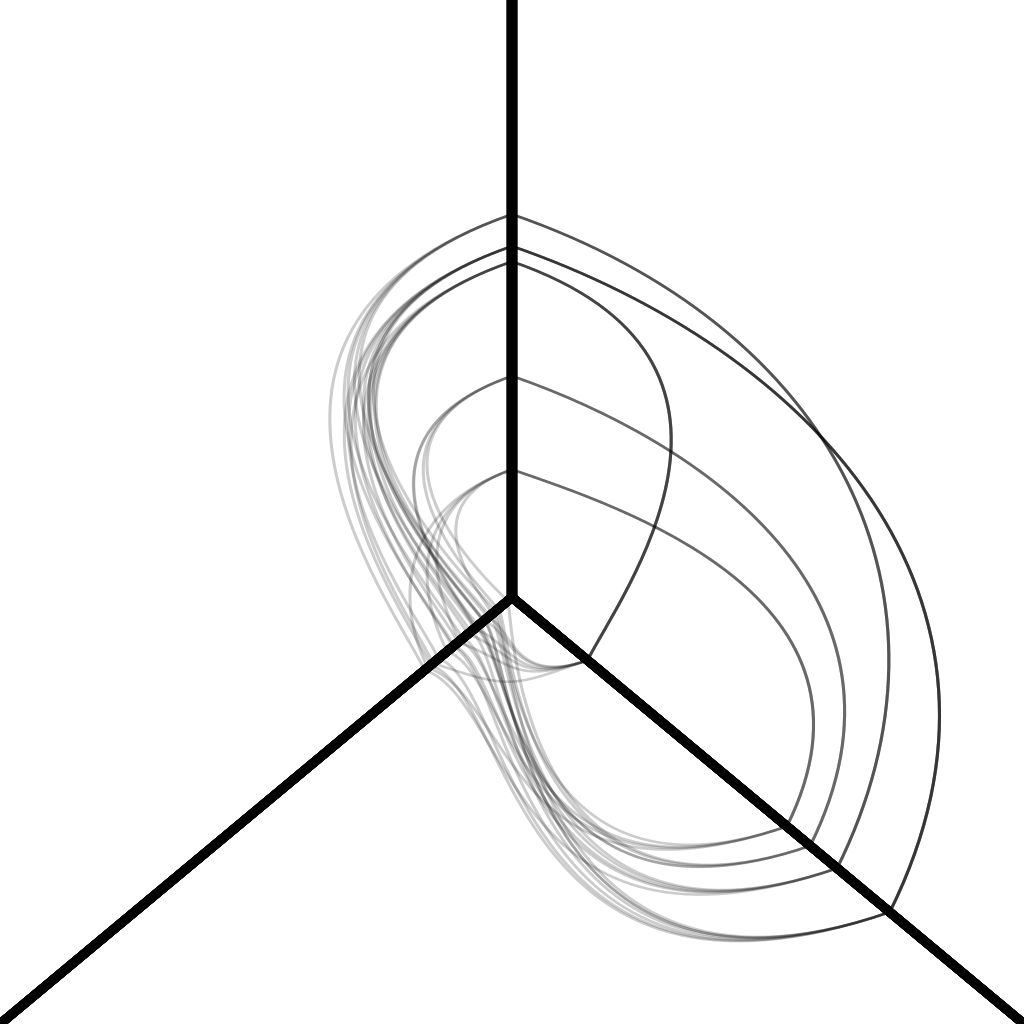} & 
         \includegraphics[width=0.24\textwidth]{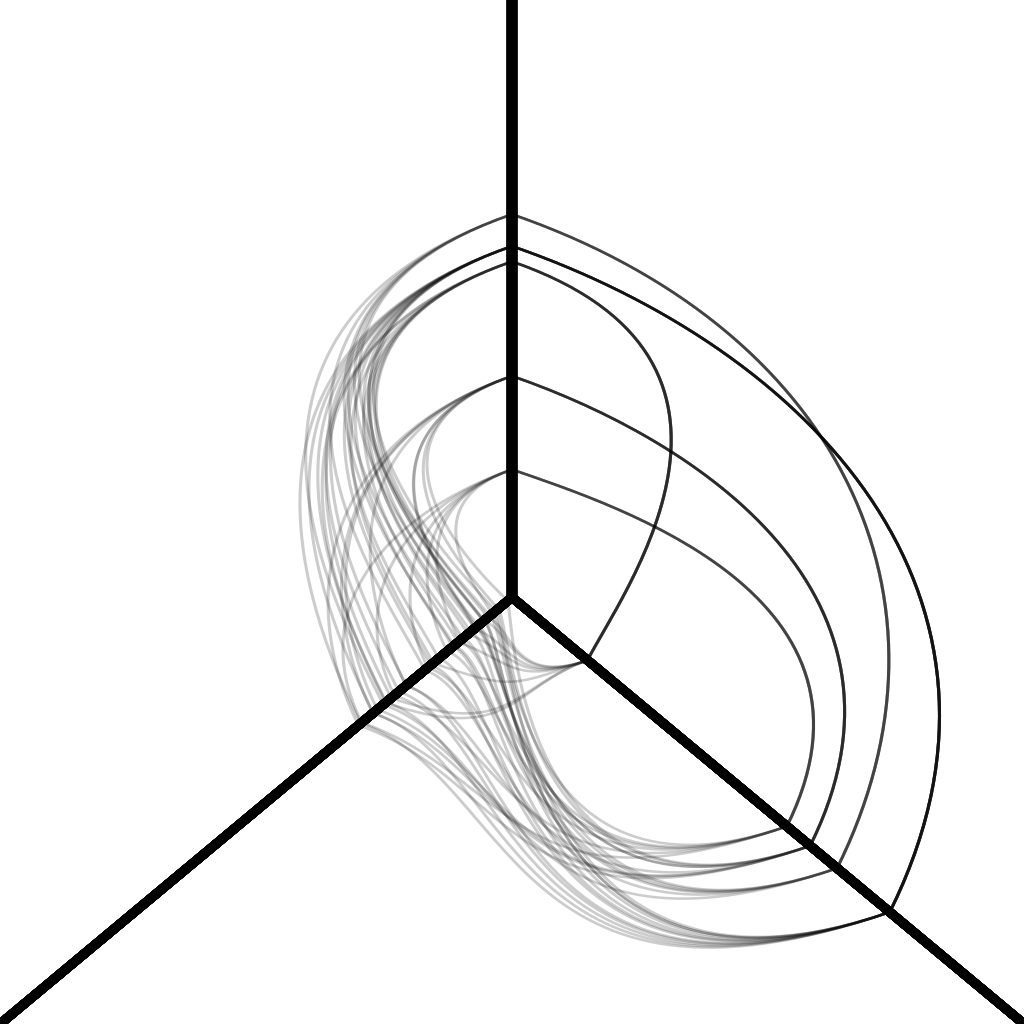} \\
         $t_4$ & $t_5$ & $t_6$ & $t_7$ \\
         \includegraphics[width=0.24\textwidth]{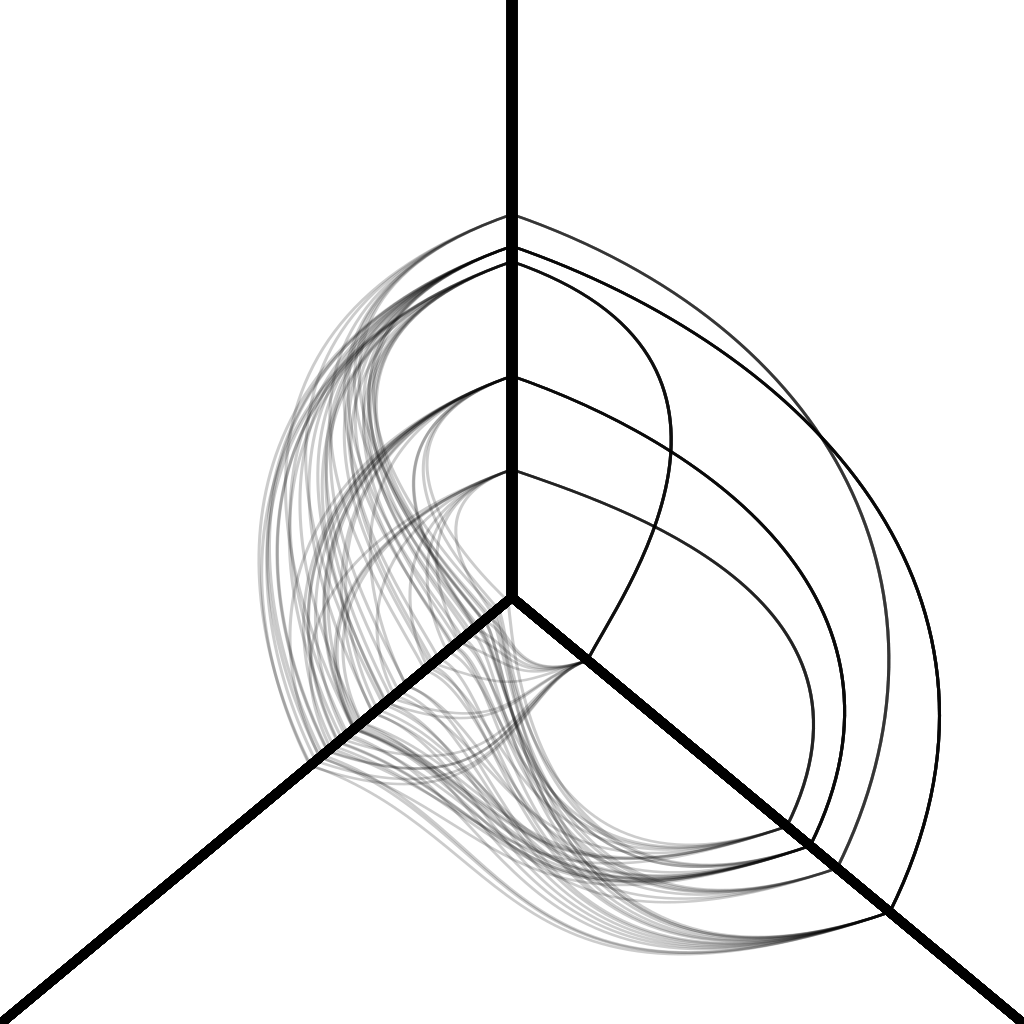} &  
         \includegraphics[width=0.24\textwidth]{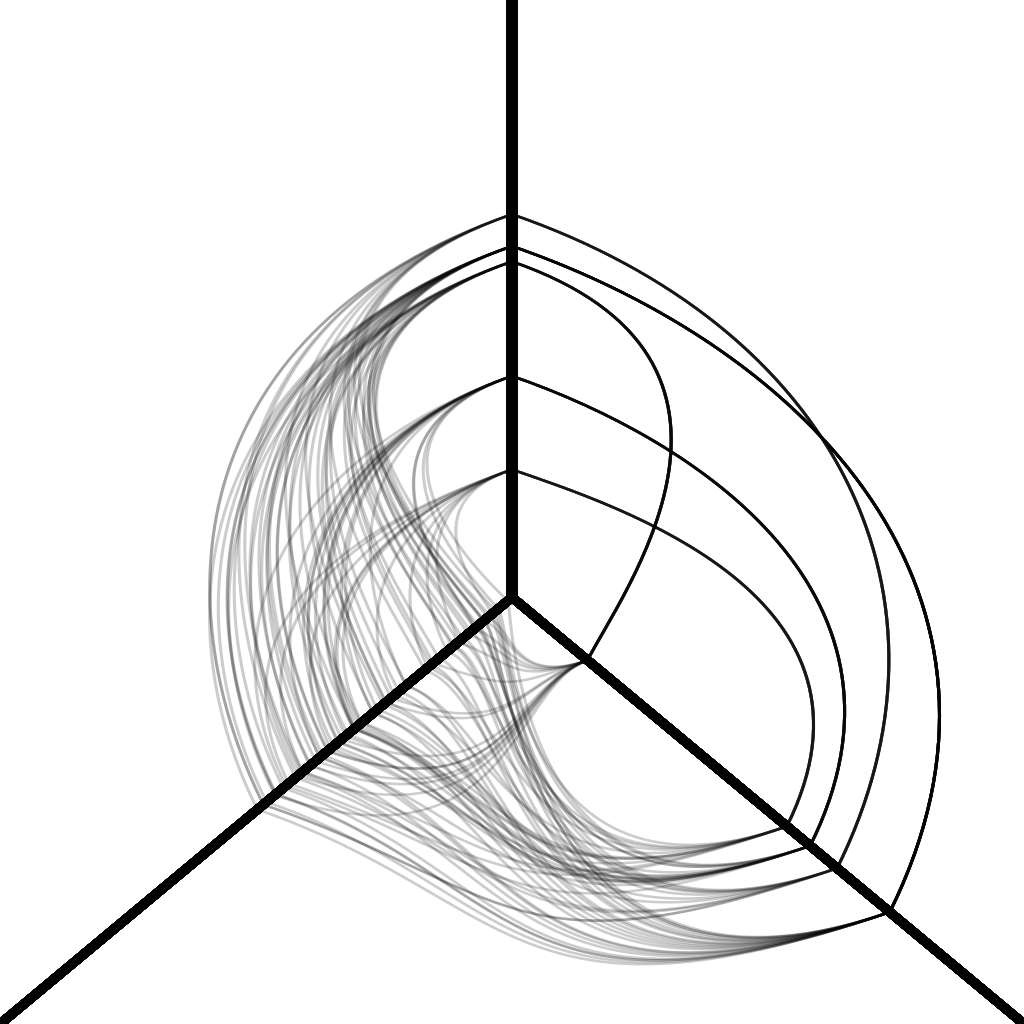} & 
         \includegraphics[width=0.24\textwidth]{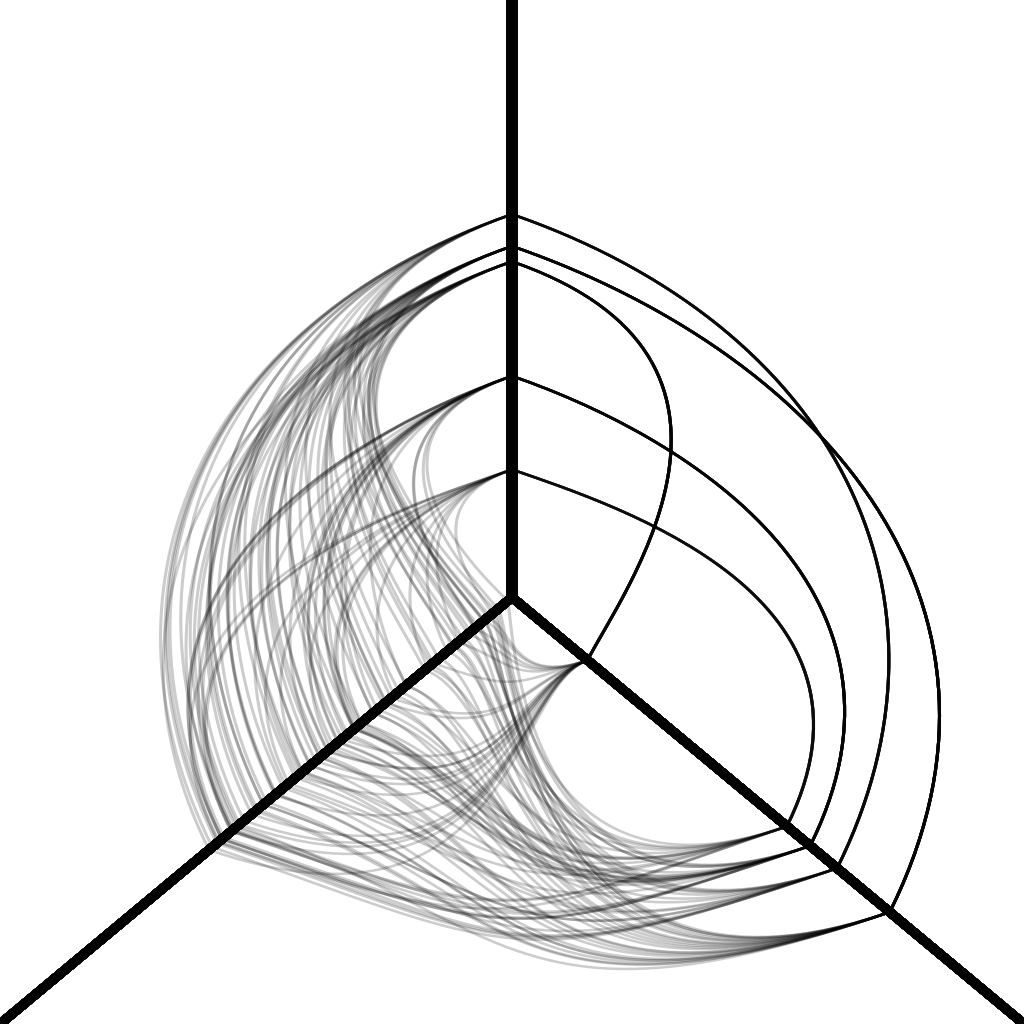} & 
         \includegraphics[width=0.24\textwidth]{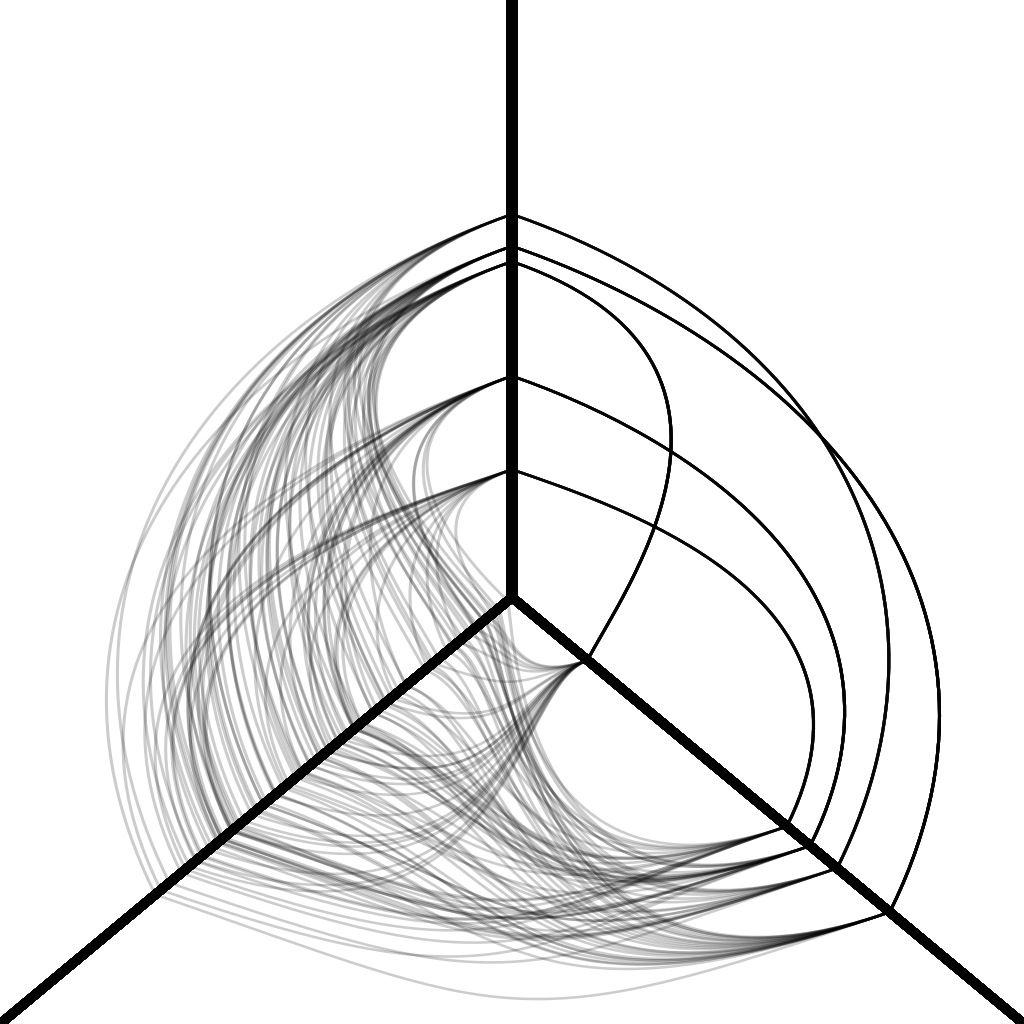} 
    \end{tabular}
    \caption{Sample time series of a DDoS attack on a honeypot target.}
    \label{fig:model}
\end{figure}

The dataset is evenly split, with 8,000 images labeled as clean traffic and 8,000 as DDoS attacks. All images are stored in \texttt{.png} format and organized into sequences of eight time-indexed frames (\(t_0\) through \(t_7\)), each capturing a snapshot of traffic at fixed intervals. The initial frame (\(t_0\)) captures the absence of traffic, and subsequent frames accumulate traffic activity over time until the sequence resets after \(t_7\). This structure allows us to treat the sequences as spatiotemporal samples for 3D convolution. Fig.~\ref{fig:model} illustrates one such 8-frame sequence depicting a DDoS event as it evolves over time.

To evaluate classifier performance across clean, augmented, and adversarial conditions, we report four standard metrics: accuracy, precision, recall, and area under the ROC curve (AUC). Each metric provides a distinct view of the model’s behavior under different testing regimes.

Accuracy (\(\mathrm{Acc}\)) measures the proportion of correctly classified samples across the dataset. It is defined as
\[
\mathrm{Acc} = \frac{1}{N} \sum_{i=1}^{N} \mathbf{1}\{\hat y_i = y_i\},
\]
where \(N\) is the total number of examples, \(y_i\) is the ground-truth label, and \(\hat y_i\) is the predicted label. While accuracy is a widely used measure, it can obscure performance in the presence of class imbalance.

Precision (\(\mathrm{Prec}\)) focuses on the correctness of positive predictions. It is computed as
\[
\mathrm{Prec} = \frac{\mathrm{TP}}{\mathrm{TP} + \mathrm{FP}},
\]
where \(\mathrm{TP}\) and \(\mathrm{FP}\) denote the number of true and false positives, respectively. High precision indicates a low false-alarm rate, which is especially important in DDoS detection where false positives can trigger unnecessary defensive actions.

Recall (\(\mathrm{Rec}\)), or sensitivity, captures the model’s ability to identify actual positives:
\[
\mathrm{Rec} = \frac{\mathrm{TP}}{\mathrm{TP} + \mathrm{FN}},
\]
with \(\mathrm{FN}\) representing false negatives. In a security context, recall is critical for minimizing missed detections of active threats.

Finally, the Area Under the ROC Curve (AUC) provides a measure of performance that is threshold-independent by evaluating the trade-off between true positive rate (TPR) and false positive rate (FPR) across all possible classification thresholds:
\[
\mathrm{AUC} = \int_{0}^{1} \mathrm{TPR}(t)\, d\bigl(\mathrm{FPR}(t)\bigr).
\]
A higher AUC indicates stronger separability between classes, independent of decision boundaries.

To assess temporal robustness and per-frame predictive power, we perform frame-wise evaluation across all test conditions: i.e., clean, augmented, FGSM-perturbed, and PGD-perturbed. Each of the eight frames in a sequence (\(t = 0, \ldots, 7\)) is treated as a stand-alone input by repeating the single image across the depth dimension to match the expected 3D CNN input shape. Perturbations are applied independently to each frame, and metrics are computed over the full validation set at each time step. This setup allows us to characterize not only the overall robustness of the model but also its ability to make accurate predictions at different stages of the observed network activity.

\subsection{Experimental details}

All experiments were conducted on an NVIDIA L40 GPU using CUDA 11.7 and PyTorch 2.0. The data preprocessing and model training code were implemented in Python 3.10. Training was performed using mixed precision with PyTorch’s automatic mixed precision module (\texttt{torch.cuda.amp}) and gradient scaling via \texttt{GradScaler}, which significantly reduced memory usage and improved runtime efficiency.

\begin{figure}[h]
    \centering
    \begin{tabular}{cccc}
         \includegraphics[width=0.24\textwidth]{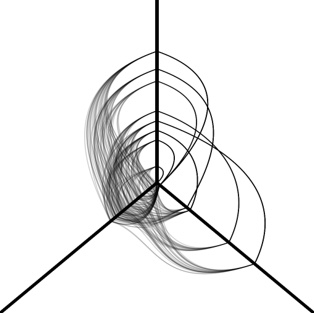} &  
         \includegraphics[width=0.24\textwidth]{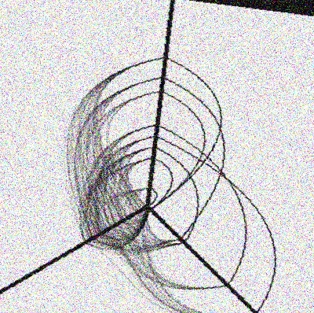} & 
         \includegraphics[width=0.24\textwidth]{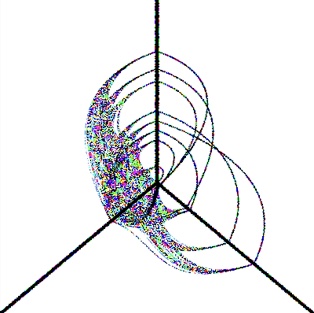} & 
         \includegraphics[width=0.24\textwidth]{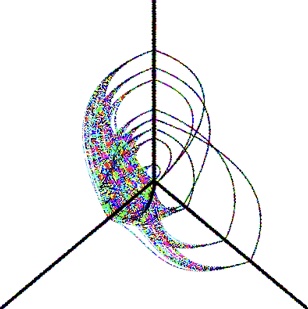} \\
         Clean & Augmented & PGD & FGSM 
    \end{tabular}
    \caption{Examples of the four input conditions used for evaluation: clean (left), augmented (center-left), and adversarially perturbed using PGD (center-right) and FGSM (right). These images illustrate the visual distortion introduced by both natural augmentations and adversarial perturbations.}
    \label{fig:regimes}
\end{figure}

Two training regimes were considered. In the clean training condition, the model was trained exclusively on unmodified, unperturbed samples. In the adversarial training condition, each minibatch of size 16 was constructed with a predefined composition designed to balance exposure to a variety of perturbations. Specifically, 8\% of the minibatch contained clean data, 12\% consisted of randomly augmented examples (with random rotation up to \(\pm18^\circ\), shear up to \(\pm11^\circ\), zoom between 0.75 and 1.0, and Gaussian noise with standard deviation \(\sigma = 0.17\)), 23\% were generated using the PGD attack with a step size \(\alpha = 1.1\), perturbation bound \(\varepsilon = 1.225\), and 40 steps, and 57\% were generated via FGSM with perturbation \(\varepsilon = 1.19\). These perturbation types and their visual characteristics are shown in Fig.~\ref{fig:regimes}.

Training used the AdamW optimizer with an initial learning rate of \(5 \times 10^{-5}\), weight decay of \(1 \times 10^{-4}\), and a batch size of 16. Learning rate scheduling followed a \texttt{ReduceLROnPlateau} policy, with a reduction factor of 0.5 and a patience of 3 epochs. Early stopping was applied after three learning rate reductions without improvement in validation metrics. Total training and evaluation runtime was approximately three hours, with no-context evaluations (frame-wise classification) taking only a few minutes.

All data were stored as preprocessed tensors in the form of \texttt{.pt} files and organized by class and split: \texttt{preprocessed\_dataset/\{train,val\}/\{Normal,DDoS\}}. Data loading was handled using PyTorch's \texttt{DataLoader} with \texttt{num\_workers=4} and \texttt{pin\_memory=True} to optimize throughput. Augmentations, when used, were applied in-batch during both the training and evaluation stages, using the same configuration across regimes to ensure consistency.

\section{Results}\label{sec:results}
We evaluate our model's performance across two training regimes, clean and adversarial, and under multiple testing conditions including clean inputs, spatial augmentations, and adversarial perturbations (PGD and FGSM). Our initial experiment involved training the model solely on clean, unperturbed sequences. Under these conditions, the model achieved perfect classification scores: 100\% accuracy, precision, and recall after just three epochs. These results suggest that the model learns to discriminate between normal and DDoS traffic effectively in the absence of noise or manipulation. However, performance degraded sharply when tested on perturbed inputs, as summarized in Table~\ref{tab:training_regime_clean}.

\begin{table}[h]
\centering
\caption{Classification performance under clean training. While clean test inputs are handled perfectly, robustness drops significantly under augmentation and adversarial attack, highlighting the model’s vulnerability when exposed to unseen perturbations.}
\begin{tabular}{llrrr}
\hline
\textbf{Training Regime} & \textbf{Condition} & \textbf{Accuracy} & \textbf{Precision} & \textbf{Recall} \\
\hline
Clean & Clean     & 1.00 & 1.00 & 1.00 \\
Clean & Augmented & 0.50 & 0.50 & 1.00 \\
Clean & PGD       & 0.55 & 0.52 & 1.00 \\
Clean & FGSM      & 0.55 & 0.52 & 1.00 \\
\hline
\end{tabular}
\label{tab:training_regime_clean}
\end{table}

To assess robustness, we introduced spatial and adversarial perturbations at inference time. As shown in Table~\ref{tab:training_regime_clean}, models trained only on clean data exhibited poor generalization to both augmented and adversarial inputs. Accuracy dropped to 50\% on spatially augmented data and hovered around 55\% on PGD- and FGSM-attacked inputs, despite recall remaining high. This gap reflects the model's inability to maintain discriminative performance under distributional shifts it was not trained to handle.

We then retrained the model under an adversarial training regime, incorporating a controlled mixture of clean, augmented, and adversarially perturbed samples in each minibatch. The retrained model demonstrated substantial gains in robustness while maintaining high performance on clean data. As shown in Table~\ref{tab:training_regime_adv}, accuracy remained above 93\% across all conditions, including under strong white-box PGD attacks. This confirms that adversarial retraining significantly improves the model’s resilience without sacrificing baseline performance.

\begin{table}[h]
\centering
\caption{Performance under adversarial training. The model generalizes well across all test conditions, including PGD and FGSM perturbations, while preserving nearly perfect accuracy on clean samples.}
\begin{tabular}{llrrr}
\hline
\textbf{Training Regime} & \textbf{Condition} & \textbf{Accuracy} & \textbf{Precision} & \textbf{Recall} \\
\hline
Adversarial & Clean     & 0.99  & 1.00  & 0.97  \\
Adversarial & Augmented & 0.985 & 0.995 & 0.96  \\
Adversarial & PGD       & 0.985 & 0.97  & 1.00  \\
Adversarial & FGSM      & 0.9325 & 0.99 & 0.855 \\
\hline
\end{tabular}
\label{tab:training_regime_adv}
\end{table}

Finally, we conducted a frame-wise evaluation to isolate the temporal contributions of each individual image within a sequence. For this no-context setup, we fed a single frame (replicated across depth) to the model without the temporal sequence. Accuracy for each frame index \(t_0\) to \(t_7\) is shown in Table~\ref{tab:timestamp_accuracy}. Predictive performance improved steadily with time, peaking between frames \(t_3\) and \(t_5\), where clean accuracy exceeded 98\% and robustness to PGD and FGSM was highest. These findings indicate that early frames are less informative due to the delayed onset of DDoS signatures, while mid-to-late frames capture stronger indicators of malicious activity.

\begin{table}[h]
\centering
\caption{Frame-wise accuracy for each timestamp under clean, augmented, and adversarial conditions. Later frames (especially \(t_3\)–\(t_5\)) are consistently more informative, suggesting they encode the clearest spatiotemporal indicators of attack behavior.}
\begin{tabular}{lrrrrrrrr}
\hline
\textbf{Condition} & $t_0$ &  $t_1$ &  $t_2$ &  $t_3$ &  $t_4$ &  $t_5$ & $t_6$ & $t_7$  \\
\hline
Clean              & 0.5000 & 0.8475 & 0.9325 & 0.9875 & 1.0000 & 0.9900 & 0.9675 & 0.9200 \\
Augmented Accuracy & 0.5000 & 0.8275 & 0.9575 & 0.9925 & 0.9575 & 0.8800 & 0.8000 & 0.7375 \\
PGD Accuracy       & 0.5000 & 0.5275 & 0.7850 & 0.9025 & 0.9150 & 0.8750 & 0.8175 & 0.7625 \\
FGSM Accuracy      & 0.5000 & 0.8300 & 0.9200 & 0.9625 & 0.9800 & 0.9875 & 0.9950 & 0.9825 \\
\hline
\end{tabular}
\label{tab:timestamp_accuracy}
\end{table}

\section{Analysis}\label{sec:analysis}

This section provides a deeper examination of the experimental findings presented in Tables~\ref{tab:training_regime_clean}–\ref{tab:timestamp_accuracy}, focusing on temporal prediction behavior, robustness benefits of adversarial training, practical latency-performance trade-offs, operational cost modeling, and remaining threat vectors in real-world deployment.

\subsection*{Temporal Prediction Dynamics}

Table~\ref{tab:timestamp_accuracy} and the associated visualizations in Fig.~\ref{fig:model} illustrate a pronounced shift in model behavior between frames \(t_2\) and \(t_4\). On clean inputs, the false positive rate drops sharply from 50\% at \(t_0\) to below 3\% by \(t_4\). Similarly, PGD-induced false negatives decline from approximately 45\% to just under 10\%. This pattern reflects the natural burst dynamics of volumetric DDoS attacks, early frames contain little activity, while later frames reveal heavy saturation across the hive-plot axes, enabling the 3D CNN to identify strong spatiotemporal patterns. These findings support a low-latency mitigation strategy: triggering an alert as soon as model confidence exceeds 0.9 by frame \(t_3\) would detect over 97\% of attacks while reducing average detection latency by nearly 60\%.

\subsection*{Robustness Improvements via Adversarial Training}

Incorporating adversarial examples during training leads to substantial robustness gains under strong attack conditions. FGSM accuracy improves from 0.55 to 0.93, and PGD accuracy from 0.55 to 0.99, with clean accuracy dropping marginally from 1.00 to 0.99. These results confirm that robustness to white-box perturbations can be improved without sacrificing baseline performance. Qualitative analysis of confusion matrices suggests that spatial augmentations (e.g., rotation, shear, zoom, noise) account for approximately 60\% of the improvement on moderately distorted samples. However, PGD-based adversarial examples play a critical role in hardening the decision boundary against stronger, iterative attacks, an effect consistent with prior work on robust image classification.

\subsection*{Latency–Accuracy Considerations}

Inference over the full 8-frame sequence requires approximately 92 milliseconds per sample on an RTX L40 GPU. In contrast, early inference using only frame \(t_3\) takes 38 milliseconds, with accuracy already exceeding 98\% at that time step. This supports the use of a sequential early-exit mechanism, in which the model returns a prediction once intermediate logits at frame \(t_3\) surpass a confidence threshold. Such a scheme nearly halves the mean inference cost with minimal reduction in detection performance. Technically, this is straightforward to implement: intermediate features from the second convolutional block can be routed through the final fully connected layer to produce provisional outputs.

\subsection*{Operational Cost Analysis}

To estimate real-world impact, we model daily cost as
\[
C = c_{\text{FP}} \cdot \mathrm{FP} + c_{\text{FN}} \cdot \mathrm{FN},
\]
where \(c_{\text{FP}} = \$0.08\) reflects the cost of throttling benign traffic and \(c_{\text{FN}} = \$12.70\) captures the estimated cost of one minute of undetected DDoS activity. Using confusion matrix statistics derived from our validation data and assuming realistic traffic patterns from the Marist honeypot, the expected daily cost drops from \$113 when using a clean-only model to just \$7.40 with the adversarially trained model, a 93\% reduction in operational expense.

\subsection*{Remaining Threat Vectors}

Despite improved robustness, the model remains exposed to a few residual attack classes. First, pixel-level adversarial perturbations that preserve overall byte-rate characteristics, such as one-pixel or C\&W-style attacks, may bypass detection. Second, adversaries may disrupt the temporal coherence of sequences by reordering frames or injecting delays, challenging the model's temporal assumptions. Third, concept drift from novel application-layer DDoS techniques could degrade accuracy over time. To mitigate these risks, we recommend pairing the model with statistical traffic monitors and scheduling periodic retraining on recent traffic snapshots to adapt to evolving patterns.

\section{Conclusion}\label{sec:conclusion}

This work extends prior research on adversarial robustness in network traffic classification by leveraging the spatiotemporal structure of hive-plot sequences from the Marist DDoS dataset. We proposed a 3D convolutional neural network capable of learning temporal dynamics from image sequences, enabling early and accurate detection of DDoS attacks. Robustness was evaluated through a combination of white-box adversarial attacks (FGSM, PGD) and stochastic input augmentations. While models trained solely on clean data performed well under benign conditions, they failed under perturbations. In contrast, models trained on a mix of clean, augmented, and adversarial data maintained high accuracy across all conditions and achieved early-stage detection performance.

These findings have practical implications for real-time intrusion detection, where latency and resilience are both critical. Adversarial retraining significantly improved model robustness without compromising clean accuracy, and accurate predictions could be made several frames prior to traffic saturation. Future work should explore generalization to unseen attack types, including black-box and adaptive adversaries, and evaluate robustness across larger, more diverse traffic datasets. Further gains may be achievable through complementary defenses such as sequence-aware input validation, continual learning, or ensembling. Optimizing the hive-plot preprocessing pipeline for real-time deployment remains an open direction for production-grade use.

\section*{Limitations}

This study has three main limitations. First, our evaluation is based on a benchmark dataset with simulated DDoS and normal traffic, which may not fully capture the diversity of real-world network behavior or unknown zero-day threats. Second, our adversarial analysis was limited to white-box attacks (FGSM, PGD) and standard augmentations; other attack classes such as black-box, C\&W, or adaptive methods were not explored. Third, the reliance on visual encoding via hive plots introduces preprocessing overhead, which, while effective for modeling, adds latency that may pose challenges in time-sensitive applications.

\begin{credits}
\subsubsection{\ackname} During the execution of this project, P.R. was funded by the National Science Foundation under grants CNS-2210091 and CNS-2136961. The authors thank the Rivas.AI Lab (\url{https://lab.rivas.ai}) for the support and helpful feedback throughout this project.

\subsubsection{\discintname}
The authors have no competing interests to declare that are
relevant to the content of this article.
\end{credits}
%
%
%
\bibliographystyle{splncs04}
\bibliography{refs}

\appendix

\section{Appendix}

This appendix summarizes key methods and mathematical used throughout the paper, including adversarial attack mechanisms and neural network operations.

\subsection*{Fast Gradient Sign Method (FGSM)}

FGSM generates adversarial inputs by computing the gradient of the loss with respect to the input and perturbing in the direction of the gradient’s sign \cite{10.48550/arxiv.2202.04821}:
$\eta = \epsilon \cdot \text{sign} \left( \nabla_x \mathcal{L}(\theta, x, y) \right)$.
Here, $\epsilon$ sets the perturbation scale, $\mathcal{L}$ is the loss, $x$ is the input, $y$ the label, and $\theta$ the model parameters.

\subsection*{Projected Gradient Descent (PGD)}

PGD extends FGSM by applying multiple iterative perturbations while projecting back into an $\epsilon$-ball around the input:
\begin{equation}
    x^{t+1} = \Pi_{\mathcal{B}_\epsilon(x)} \left( x^t + \alpha \cdot \text{sign} \left( \nabla_x \mathcal{L}(\theta, x^t, y) \right) \right),
    \label{eq:pgd}
\end{equation}
where $\Pi_{\mathcal{B}_\epsilon(x)}$ denotes projection, and $\alpha$ is the step size.

\subsection*{Global Average Pooling (3D)}

3D global average pooling compresses each feature map to a scalar by averaging over depth ($D$), height ($H$), and width ($W$):
\begin{equation}
g_c = \frac{1}{D\,H\,W}
\sum_{d=1}^{D}\sum_{h=1}^{H}\sum_{w=1}^{W}
X_{c,\,d,\,h,\,w}
\label{eq:gap3d}
\end{equation}

\subsection*{3D Convolution}

3D convolution extends the 2D operation to include depth, enabling the model to learn spatiotemporal features from input tensors:
\begin{equation}
Y_{c',\,d,\,h,\,w}
= \sum_{c=1}^{C_{\text{in}}}
\sum_{u=1}^{k_d}\sum_{v=1}^{k_h}\sum_{w'=1}^{k_w}
K_{c',c,u,v,w'}\;
X_{c,\,d+u-1,\,h+v-1,\,w+w'-1}
\label{eq:conv3d}
\end{equation}
This operation forms the core of video and sequence-based CNNs as originally proposed in \cite{10.1109/iccv.2015.510}.

\end{document}